\documentclass[12pt]{article}
\usepackage{epsfig}
\usepackage{amssymb}
\usepackage{amsmath}

\begin{document}

  \title{\LARGE \bf  Boundary Conditions and Modes\\ of the Vertically Hanging Chain  \vskip 1.1cm }

   \author{
  \large  Y. Verbin
  \thanks{{Electronic addresses: verbin@openu.ac.il} } }
 \date{ }
   \maketitle
    \centerline{ \em Department of Natural Sciences, The Open University
   of Israel}
   \centerline{\em Raanana 43107, Israel}
   \vskip 1.1cm

 The hanging chain is a very instructive system for demonstrating more advanced methods and ideas for the analysis of normal modes of one-dimensional systems, beyond the standard ordinary (horizontal) string. Accordingly, the normal modes of the hanging chain are analyzed with several cases of boundary conditions and are compared to the ordinary vibrating string.

\newpage
\setcounter{page}{1}

\section{Introduction}\label{Int}
\setcounter{equation}{0}

Transverse waves in the hanging chain are frequently studied in texts of classical mechanics or wave theory  \cite{Lamb1929,Routh1955} and in several papers, usually generalizing the system in various directions \cite{Satterly,McCreeshEtAl,Levinson,Western,SilvermanEtAl,Bailey,NoelEtAl}. The basic situation is of a (continuous) chain with an upper fixed end and a lower free one and the normal modes and the corresponding angular frequencies (which we will call simply ``frequencies'') are the main subject of interest.

This is actually a special case of the more general situation described by the one-dimensional wave equation for a string with a non-uniform mass density (per unit length) $\rho (x)$  and a tension  $T(x)$:
\begin{equation}
\rho (x)\frac{\partial^2 \psi}{\partial t^2}-\frac{\partial }{\partial x}\left( T(x)\frac{\partial \psi}{\partial x} \right )=0
\label{WaveEqTvar} .
\end{equation}

The case of the hanging uniform chain reduces to the wave equation with a constant mass density   and a linearly increasing tension $T(x)=\rho gx$  with the additional conditions that the chain is free at its lower end $x=0$ and fixed at its upper end $x=L$.

More explicitly, the wave equation for this case becomes
\begin{equation}
\frac{\partial^2 \psi}{\partial t^2}- gx \frac{\partial^2 \psi}{\partial x^2}
-g\frac{\partial \psi}{\partial x}=0
\label{WaveEqVerCh}
\end{equation}
and the normal modes $u(x)$ defined by $\psi = u(x)\cos(\omega t)$ solve the equation
\begin{equation}
 gx \frac{\partial^2 u}{\partial x^2}
+g\frac{\partial u}{\partial x}+\omega^2 u=0
\label{VerChModes} .
\end{equation}
The usual procedure is to transform this mode equation to Bessel equation with index 0
\begin{equation}
z^2 u'' + z u' +z^2 u=0
\label{Bessel0}
\end{equation}	
by $x=gz^2/4\omega^2$  and to argue that since $Y_0 (z)$  is singular at $z=0$, the normal modes are given by $J_0 (z)$ only.

From a pedagogical point of view, all these discussions are lacking a decent comparison of the situation with the more well-known situation of standing waves in an ordinary (constant tension $T$) string with either free or fixed boundary conditions. The normal frequencies in the latter situation result from imposing boundary conditions (BC) at both ends. E.g: imposing $u(0)=u(L)=0$ yields the following set of frequencies:
\begin{equation}
\omega_n = \sqrt{\frac{T}{\rho}}\frac{\pi}{L}n \hspace {0.2cm},\hspace {0.4cm} n=1,2,3...
\label{FreqFixed}
\end{equation}
while imposing $u'(0)= u(L)=0$ yields a different set:
\begin{equation}
\omega_n = \sqrt{\frac{T}{\rho}}\frac{\pi}{L}\left(n-\frac{1}{2}\right) \hspace {0.2cm},\hspace {0.4cm} n=1,2,3...
\label{FreqFree} .
\end{equation}
The situation of the hanging chain is different in that the normal frequencies result from a single boundary condition at $x=L$  (or $z= \bar{z}\equiv 2\omega \sqrt{L/g}$) only. Actually this is not completely true, since the information about the other end is hidden in the decision to discard the $Y_0 (z)$ contribution.

Anyhow,  we were unable to find any analogous discussion about the other BC for the chain, namely the case of a fixed lower end. This is the direction we intend to proceed in this note: analyze the normal modes of the hanging chain and some simple generalization which we present in order to get a wider framework for understanding this system.

\section{Normal Modes of the Hanging Chain}\label{Modes}
\setcounter{equation}{0} \setcounter{table}{0} \setcounter{figure}{0}

\subsection{The Simple Hanging Chain}\label{SC}

\begin{figure}[b]
\hbox to\linewidth{\hss%
    \resizebox{8cm}{5cm}{ \includegraphics{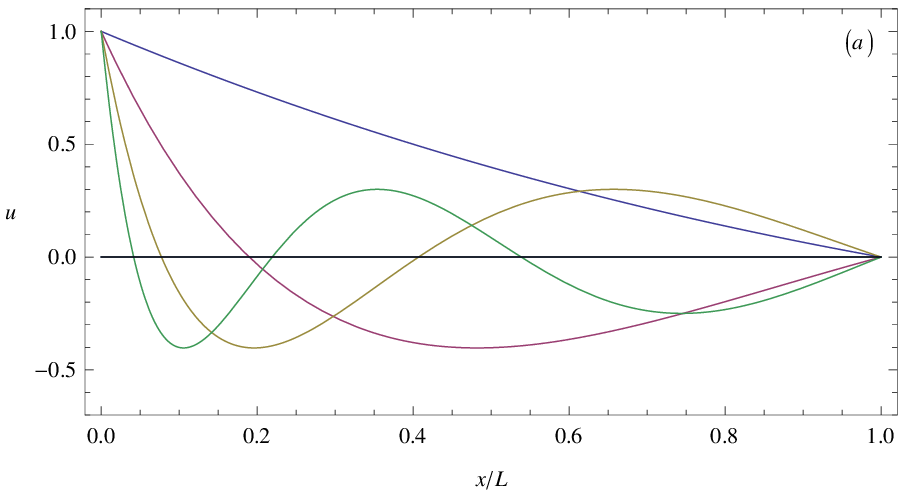}}
\hspace{5mm}%
        \resizebox{8cm}{5cm}{\includegraphics{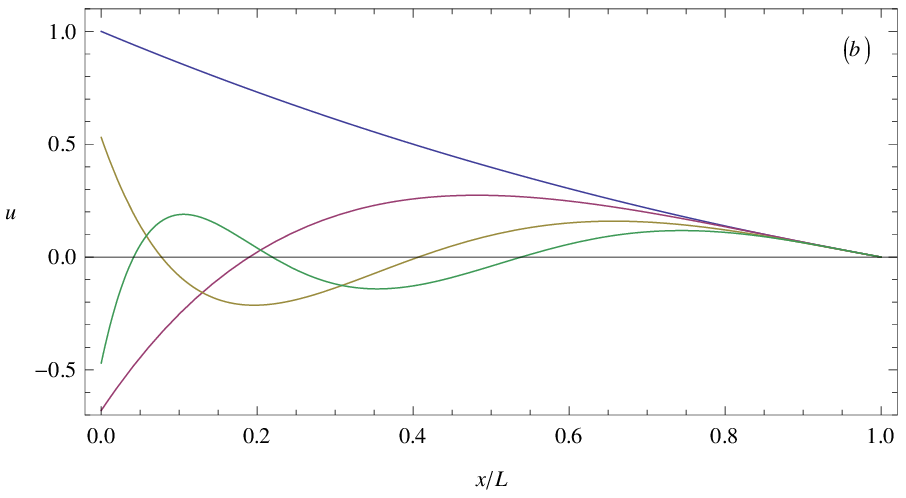}}
\hss}
\caption{\small {
The first 4 modes of the hanging chain. The scale is chosen such that all modes have: (a)  the same maximum displacement; (b)  the same slope at $x=L$.}}
\label{HangCModes}
\end{figure}

First we obtain the modes of the hanging chain with a lower free end and an upper fixed one. As mentioned above, Eq. (\ref{Bessel0}) is a Bessel equation with index 0, so its general solution is given by the combination of the two independent Bessel and Neumann function with index 0
\begin{equation}
u(z) = a J_0 (z) + b Y_0 (z)
\label{uCombin}.
\end{equation}
Since $ Y_0 (z)$  is singular at the lower end, it cannot contribute and $b=0$. The natural frequencies are therefore obtained by solving the condition obtained from imposing an upper fixed end:
\begin{equation}
J_0 (\bar{z}) =0
\label{BesselZeroes}.
\end{equation}
The zeroes of the Bessel functions are tabulated in numerous sources and mathematical packages. So if we designate the zeroes of $J_0 (z)$  by $\zeta_n$ , we can get the following expression for the natural frequencies of this system:
\begin{equation}
\omega_n = \frac{1}{2}\sqrt{\frac{g}{L}}\zeta_n = \frac{\Omega}{2}\zeta_n
\label{FreqHang}
\end{equation}
where we notice the appearance of the simple pendulum frequency $\Omega =\sqrt{g/L}$. The corresponding modes are therefore
\begin{equation}
u_n = aJ_0(\zeta_n\sqrt{x/L}) \hspace {0.2cm},\hspace {0.4cm} 0\leq x \leq L
\label{ModesHang}
\end{equation}
Table \ref{TBesselZeroes} gives the first 6 zeroes and corresponding frequencies of the hanging chain.
\begin{table}
  \centering
\begin{tabular}{@{}ccccccc@{}}
\hline \hline
$n$   &1         &2            &3          &4                  &5                  &6\\
\hline
$\zeta_n$   &2.40483           &5.52008        &8.65373          &11.79153                   &14.93092                 &18.07106  \\
\hline
$\omega_n/\Omega$    &1.20241
&2.76004        &4.32686          &  5.89577                   &  7.46546                 &  9.03553  \\
\hline \hline
\end{tabular}
  \caption{the first 6 zeroes and corresponding frequencies of the hanging chain}\label{TBesselZeroes}
\end{table}
Fig. \ref{HangCModes} presents the first 4 modes of the hanging chain. All that is well-known for many years \cite{Lamb1929,Routh1955}.

Now we move to the possibility of adding a boundary condition at the lower end. The simplest possibility is a fixed lower end. If we insist on imposing a second boundary condition $u(0)=0$ , it is obvious that the only solution is $u(x)=0$ . That is, there exist no eigenmodes of the hanging chain with both ends fixed. Similarly, there exist no solutions with the second popular kind of boundary condition $u'(0)=0$.

We can however modify the system slightly in order to get some insight as to the physical reasons behind this mathematical facts.

\subsection{The Hanging Chain with a Bead}\label{SCB}

The simplest modification is to add a small bead of mass $m$ at the lower end of the string. The string tension in this case is   $T(x)=mg+\rho gx$ so Eq. (\ref{VerChModes}) changes into
\begin{equation}
(\mu +x) g \frac{\partial^2 u}{\partial x^2}
+g\frac{\partial u}{\partial x}+\omega^2 u=0
\label{VerChModesBead}
\end{equation}	
where  $\mu=m/\rho$. This is the same as Eq. (\ref{VerChModes}) up to a simple translation and therefore can be transformed similarly to the same Bessel equation (\ref{Bessel0}) by $x=gz^2/4\omega^2 - \mu$. The general solution of Eq. (\ref{Bessel0}) is $u(z) = a J_0 (z) + b Y_0 (z)$ as before, but now the lower point $x=0$ does not correspond to $z=0$ , but to
$z=z_{-}\equiv 2\omega \sqrt{\mu/g}$ , so  $Y_0 (z)$ is not excluded from the solutions. Note that the upper end corresponds now to  $z=z_{+}\equiv 2\omega \sqrt{(L+\mu)/g}$.
So in order to obtain the modes in this modified case we have to impose the fixed end boundary conditions in exact analogy with the simple string case, at both ends:
\begin{equation}
a J_0 (z_{-}) + b Y_0 (z_{-})=0 \hspace {0.4cm},\hspace {0.4cm} a J_0 (z_{+}) + b Y_0 (z_{+})=0
\label{BCHangBead}.
\end{equation}
These are two linear homogeneous equations for the coefficients $a$,$b$ which have non-trivial solution, as is well known, if the corresponding determinant vanishes:
\begin{equation}
J_0 (z_{-})Y_0 (z_{+}) - J_0 (z_{+})Y_0 (z_{-})=0
\label{DetEqZero}.
\end{equation}
For a given $L$ and $m$, this is an ``algebraic'' equation for the frequency $\omega$  which gets the explicit form
\begin{eqnarray} \nonumber
J_0 (2 \sqrt{\mu/L} \hspace {0.2cm} \omega/\Omega)Y_0 (2 \sqrt{1+\mu/L} \hspace {0.2cm} \omega/\Omega) - \hspace {5cm}\\
 J_0 (2 \sqrt{1+\mu/L} \hspace {0.2cm} \omega/\Omega)Y_0 (2 \sqrt{\mu/L} \hspace {0.2cm} \omega/\Omega)=0
\label{FreqHangBead}.
\end{eqnarray}
Solving this equation can only be done numerically, although in a straightforward manner. Let us define
\begin{equation}
f(\bar{\mu},\bar{\omega})=J_0 (2\sqrt{\bar{\mu}}\hspace {0.2cm}\bar{\omega})Y_0 (2\sqrt{1+\bar{\mu}}\hspace {0.2cm}\bar{\omega}) - J_0 (2\sqrt{1+\bar{\mu}}\hspace {0.2cm}\bar{\omega})Y_0 (2\sqrt{\bar{\mu}}\hspace {0.2cm}\bar{\omega})
\label{FreqHangBeadfctn}
\end{equation}
and explore its zeroes in the rescaled $\bar{\mu}$, $\bar{\omega}$  plane. Although it is impossible to cover numerically the entire ($\bar{\mu}$, $\bar{\omega}$) plane, mapping a sufficiently large portion of it will provide us with a good understanding and quantitative results.

\begin{figure}[t]
\hbox to\linewidth{\hss%
    \resizebox{8cm}{7.1cm}{ \includegraphics{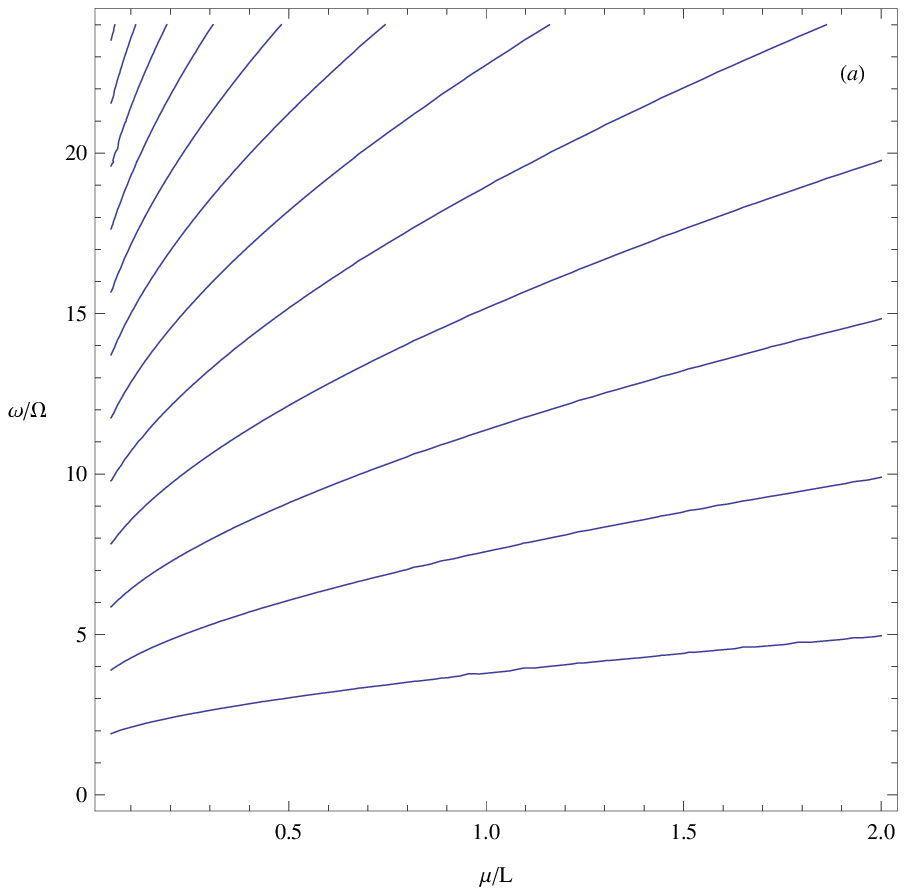}}
\hspace{5mm}%
        \resizebox{8cm}{7.1cm}{\includegraphics{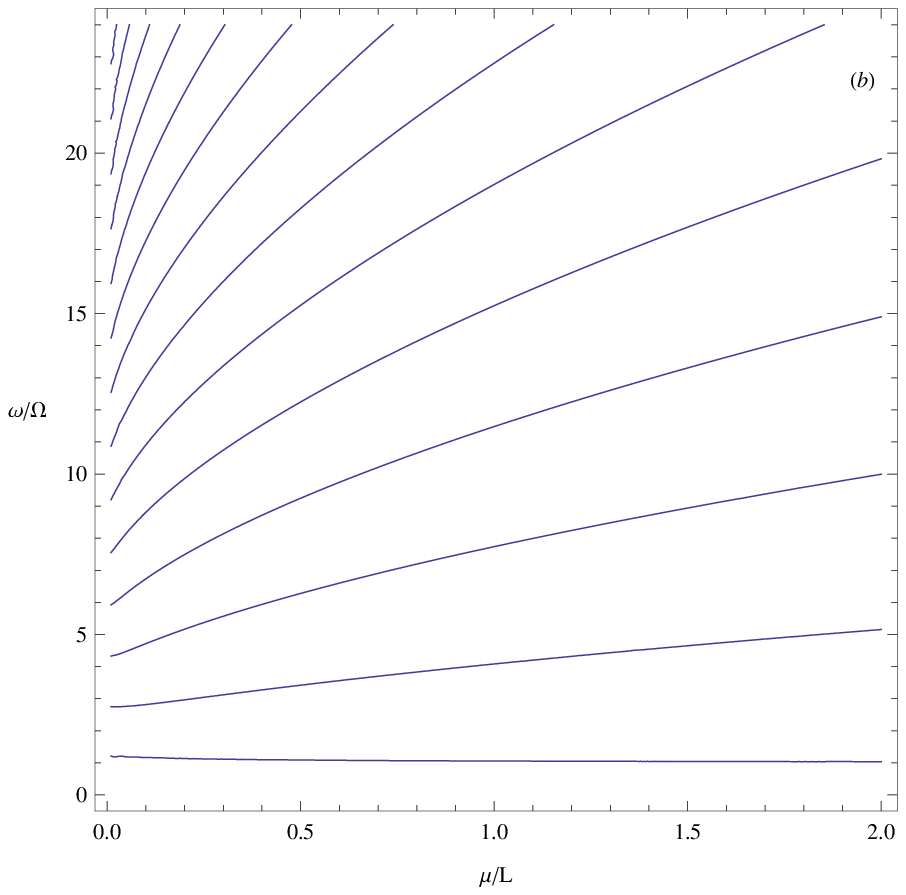}}
\hss}
\caption{\small {
The lower frequencies $\bar{\omega}=\omega/\Omega$ as a function of $\bar{\mu}=m/\rho L$ for a hanging chain with a bead: (a) fixed lower end; (b) free lower end - for explanation see towards the end of the section.}}
\label{HangCFreqs}
\end{figure}

Fig. \ref{HangCFreqs}(a) depicts the level curves of $f(\bar{\mu},\bar{\omega})=0$ for the region $0<\bar{\mu}\leq 2$, $0< \bar{\omega}\leq 24$- that is the lower frequencies as a function of $\bar{\mu}$. Part (b) of the figure is explained in what follows.  Note the different ``quantization'' patterns in the two cases.


As a simple check we notice that for large $\bar{\mu}$  the frequencies approach the simple string ones
\begin{equation}
\omega_n = \sqrt{\frac{mg}{\rho}}\frac{\pi}{L}n = \sqrt{\frac{m}{\rho L}}\Omega\pi n \hspace {0.2cm},\hspace {0.4cm} n=1,2,3...
\label{FreqFixedLimit}
\end{equation}
obtained from Eq. (\ref{FreqFixed}) since the tension is almost constant and dominated by the bead mass $m$.

The way to get the frequencies in this limit is to use the well-known asymptotic expression for the Bessel and Neumann function:
\begin{eqnarray} \nonumber
J_\nu (z) \sim \sqrt{\frac{2}{\pi z}} \cos(z-\frac{n\pi}{2} -\frac{\pi}{4}) \\
Y_\nu (z) \sim \sqrt{\frac{2}{\pi z}} \sin(z-\frac{n\pi}{2} -\frac{\pi}{4})
\label{AsympBessel}.
\end{eqnarray}
Using this approximation in Eq. (\ref{DetEqZero}) gives
\begin{equation}
\frac{2}{\pi}\frac{1}{\sqrt{z_{+}z_{-}}}\sin(z_{+}-z_{-})=0
\label{DetAsympt}
\end{equation}
or $z_{+}-z_{-} = n\pi$. Writing  $z_{+}$ and $z_{-}$ in terms of the physical parameters and expanding in terms of $\bar{\mu}$ for $\bar{\mu}=\mu/L=m/\rho L >>1$ gives up to second order
\begin{equation}
\frac{\omega_n}{\Omega}\sqrt{\frac{\rho L}{m}}\left(1-\frac{1}{4}\frac{\rho L}{m}\right )=n\pi
\label{FreqAsympt}.
\end{equation}
If we just ignore the second term in the brackets, we return to the frequencies of Eq. (\ref{FreqFixedLimit}) which are valid for a bead which is much massive than the string (chain), such that the string tension is uniform (compare also with Eq. (\ref{FreqFixed})). The lowest order correction to the frequencies of Eq. (\ref{FreqFixedLimit}) is obtained by the following expansion up to first order:
\begin{equation}
\omega_n =\sqrt{\frac{mg}{\rho L^2}}\frac{n\pi}{1-\frac{1}{4}\frac{\rho L}{m}}\approx
\frac{n\pi}{L}\sqrt{\frac{mg}{\rho}}\left(1+\frac{1}{4}\frac{\rho L}{m}\right)
\label{FreqAsympt2}.
\end{equation}
This result contains the uniform tension contribution of (\ref{FreqFixedLimit}) and an additional one of the order of the ratio between the string mass $\rho L$ and the bead mass $m$.

\begin{figure}[t]
\centering
\includegraphics{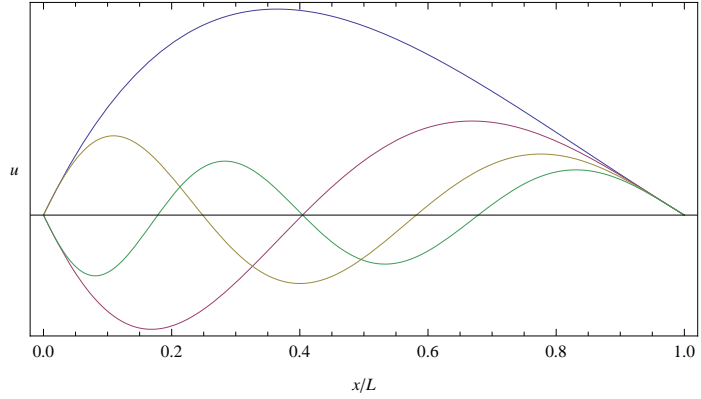}
\caption{\small {The first 4 modes of the hanging chain with a bead of $\bar{\mu}=0.25$. Fixed lower end.}}
\label{HangC+MassModes}
\end{figure}

A typical situation of the first 4 modes at an intermediate case where $\bar{\mu}=0.25$ is shown in Fig. \ref{HangC+MassModes}. The corresponding frequencies are obtained by finding numerically the roots of $f(\bar{\mu},\bar{\omega})$ of Eq. (\ref{FreqHangBeadfctn}) - see also Fig. \ref{HangCFreqs}(a). The values of the first 4 ones are $\omega/\Omega=$ $2.52155, \,  5.07251, \,  7.61757, \, $ $10.16095$. Note the approximate harmonic law $\omega \approx \omega_1 n$.

\begin{figure}[b]
\centering
\includegraphics{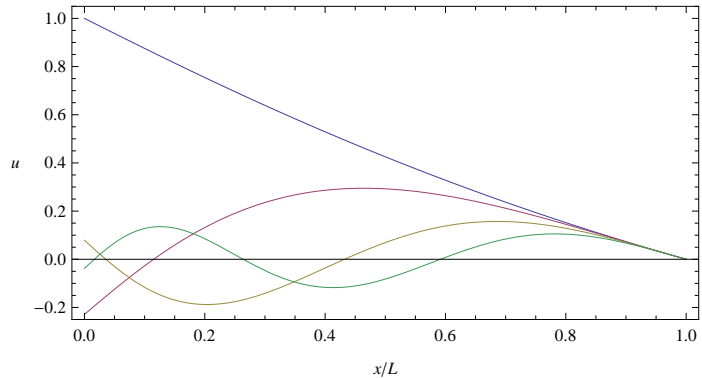}
\caption{\small {The first 4 modes of the hanging chain with a bead of $\bar{\mu}=0.25$. Free lower end.}}
\label{HangC+MassModesFree}
\end{figure}

It is obvious that due to the different set of BC, the limit $m\rightarrow 0$ will not reproduce the free case frequencies of Eq. (\ref{FreqHang}). These are obtained as the limit of the frequencies for the modes of the free lower end. The corresponding BC in this case is
\begin{equation}
\left [m\frac{\partial^2 \psi}{\partial t^2}-T(x)\frac{\partial \psi}{\partial x}\right]_{_{x=0}}=0
\label{FreeBCHangC+M}
\end{equation}
which in terms of $u(x)$ reads $\omega^2 u(0)+gu'(0)=0$ and becomes the following condition on $u(z)$:
\begin{equation}
2u'(z_{-})+z_{-} u(z_{-})=0
\label{FreeBCHangC+M-u}.
\end{equation}
The modes of the free lower end case are therefore still given by  the combination (\ref{uCombin}), but now with the following set of conditions instead of Eq. (\ref{BCHangBead}):
\begin{equation}
a (z_{-}J_0 (z_{-})+2 J'_0 (z_{-}))+ b (z_{-}Y_0 (z_{-})+2 Y'_0 (z_{-}))=0 \hspace {0.4cm},\hspace {0.4cm} a J_0 (z_{+}) + b Y_0 (z_{+})=0
\label{FreeBCHangBead}.
\end{equation}

We may simplify matters a little using the identity $J'_0 (z)=-J_1 (z)$ and similarly $Y'_0 (z)=-Y_1 (z)$, and get the frequencies from the zeroes of
\begin{eqnarray} \nonumber
g(\bar{\mu},\bar{\omega})=[\sqrt{\bar{\mu}}\hspace {0.2cm}\bar{\omega}J_0 (2\sqrt{\bar{\mu}}\hspace {0.2cm}\bar{\omega})-J_1 (2\sqrt{\bar{\mu}}\hspace {0.2cm}\bar{\omega})]Y_0 (2\sqrt{1+\bar{\mu}}\hspace {0.2cm}\bar{\omega}) \hspace {2cm} \\
-[\sqrt{\bar{\mu}}\hspace {0.2cm}\bar{\omega}Y_0 (2\sqrt{\bar{\mu}}\hspace {0.2cm}\bar{\omega})-Y_1 (2\sqrt{\bar{\mu}}\hspace {0.2cm}\bar{\omega}) ] J_0 (2\sqrt{1+\bar{\mu}}\hspace {0.2cm}\bar{\omega})
\label{FreqHangBeadfctnFree}.
\end{eqnarray}
The level curves of $g(\bar{\mu},\bar{\omega})=0$ for the region $0<\bar{\mu}\leq 2$, $0< \bar{\omega}\leq 24$ - that is the lower frequencies as a function of $\bar{\mu}$  are shown in Fig. \ref{HangCFreqs}(b). Figure \ref{HangC+MassModesFree} presents the first 4 modes for $\bar{\mu}=0.25$ and a free lower end. The corresponding frequencies are $\omega/\Omega=$ $1.11975, \,  3.04202, \,  5.37191, \, 7.82383$.

The $\mu=0$ values of the curves in Fig. \ref{HangCFreqs}(b) indeed reproduce the frequencies of Table \ref{TBesselZeroes}. In order to see this analytically, we should take the limit $\bar{\mu}\rightarrow 0$ in Eq. (\ref{FreqHangBeadfctnFree}). The function $g(\bar{\mu},\bar{\omega})$ reduces in this limit to the product
$h(\bar{\omega})J_0 (2\bar{\omega})$ where
\begin{equation}
h(\bar{\omega})=\lim_{\bar{\mu}\rightarrow 0} [Y_1 (2\sqrt{\bar{\mu}}\hspace {0.2cm}\bar{\omega}) - \sqrt{\bar{\mu}}\hspace {0.2cm}\bar{\omega}Y_0 (2\sqrt{\bar{\mu}}\hspace {0.2cm}\bar{\omega})]
\label{h-fctn}.
\end{equation}
This function is not a constant, but it is negative definite for all $0<2\sqrt{\bar{\mu}}\hspace {0.2cm}\bar{\omega}<3.38424$, so for $\bar{\mu}\rightarrow 0$ the prefactor of $h(\bar{\omega})J_0 (2\bar{\omega})$ becomes negative for all $\bar{\omega}$ and the zeroes become those of
$J_0 (2\bar{\omega})$. We thus recover the characteristics (\ref{BesselZeroes})-(\ref{FreqHang}) for the simple hanging chain.

Note also that the BC on $u(x)$ at the lower end, $\omega^2 u(0)+gu'(0)=0$ holds also in the $m=0$ case and actually follows from the mode equation (\ref{VerChModes}) for $x=0$. The lower end is therefore not entirely free; its slope and amplitude are related in a specific way which also involves the frequency.

\subsection{The Hanging Chain Held Fixed at a Point above the Lower End}\label{SCC}

A second possibility to modify the fixed lower end BC is to choose the lowest node to be somewhat higher above the lowest end, i.e. at $x=x_0 > 0$. This is a non-singular point of Eq. (\ref{VerChModes}) and now solutions satisfying $u(x_0)=u(L)=0$ do exist. Actually, we do not have to solve the equation again, but we may rather use the previous solutions for the chain with the bead, since there is an obvious equivalence between the two systems: The simple chain with $u(x_0)=u(L)=0$ is the same as the chain with a bead whose length is $L-x_0$ and the bead mass is $m=\rho x_0$, or $\mu = x_0$.

\section{Conclusion}\label{Conc}
\setcounter{equation}{0}

We have seen that the hanging chain is a very instructive system in order to demonstrate the slightly more advanced methods and ideas used in order to analyze the normal modes of one-dimensional systems. The too simple and too special cases of the ordinary vibrating string are fine for a basic introduction of the matters, but may mislead the student unless an explicit comparison with more ``interesting'' systems like the hanging chain is carried out.

{\bf acknowledgments}

The author thanks Prof. Aharon Davidson for useful discussions and helpful suggestions.



\begin{thebibliography}{99}

   \bibitem{Lamb1929}
  H.~Lamb,
  \textit{ Higher Mechanics },
  (Cambridge University Press, Cambridge 1929), pp. 225-226.
 \bibitem{Routh1955}
  E.J.~Routh,
  \textit{The Advanced Part of A Treatise on the Dynamics of a System of Rigid Bodies},
  (Dover Publications, New York 1955), pp. 404-405.
  \bibitem{Satterly}
  J.~Satterly, ``Some Experiments in Dynamics, Chiefly on Vibrations'', Am.\ J.\ Phys.\  {\bf 18}, 405-416 (1950).
  \bibitem{McCreeshEtAl}
  J.P.~McCreesh et al, ``Vibrations of a hanging chain of discrete links'', Am.\ J.\ Phys.\  {\bf 43}, 646-648 (1975).
  \bibitem{Levinson}
  D.A. ~Levinson, ``Natural frequencies of a hanging chain '', Am.\ J.\ Phys.\  {\bf 45}, 680-681 (1977).
  \bibitem{Western}
  A.B.~Western, ``Demonstration for observing $J_0 (x)$ on a resonant rotating vertical chain'', Am.\ J.\ Phys.\  {\bf 48}, 54-56 (1980).
  \bibitem{SilvermanEtAl}
  M.P.~Silverman et al, ```String theory': equilibrium configurations of a helicoseir '', Eur.\ J.\ Phys.\  {\bf 19}, 379-387 (1998).
  \bibitem{Bailey}
  H.~Bailey, ``Motion of a hanging chain after the free end is given an initial velocity'', Am.\ J.\ Phys.\  {\bf 68}, 764-767 (2000).
  \bibitem{NoelEtAl}
  J-M.~Noel et al, ``Natural configurations and normal frequencies of a vertically suspended, spinning, loaded cable with both extremities pinned'', Eur.\ J.\ Phys.\  {\bf 29}, N47-N53 (2008).

\end{thebibliography}
\end{document}